\documentclass[aps,prd,preprint,nofootinbib,showpacs,letterpaper]{revtex4-1}
\usepackage{amsmath,amssymb,amsfonts}
\usepackage{slashed}
\allowbreak
\allowdisplaybreaks
\usepackage{graphicx}
\usepackage{hyperref} 

\begin{document}
\title{$D=2$ gluon condensate and QCD propagators at finite temperature}

\author{Purnendu Chakraborty}
\affiliation{Physical Research Laboratory, Navrangpura, Ahmedabad 380009,
 Gujarat, India.}

\author{Munshi G. Mustafa}
\affiliation{Saha Institute of Nuclear Physics, 1/AF Bidhannagar, 
Kolkata 700064, India}


\pacs{11.10.Wx,12.38.Lg,12.38.Mh,25.75.Nq}
\begin{abstract}
We calculate the dimension two gluon condensate contribution to quark, 
gluon and ghost propagators at finite temperature.
\end{abstract}
\maketitle

Quantum Chromodynamics (QCD) has been firmly established as the correct theory 
of strong interaction. At high enough temperature, the QCD 
matter undergoes a rapid crossover~\cite{Aoki:2006we}from confined hadronic 
phase to a  
quasifree gas of deconfined quarks and gluons called quark 
gluon plasma~\cite{kapusta2006finite}.  The structure of QCD near the phase 
transition seems to be rather complex and dominated by strong non-perturbative 
effects~\cite{Boyd:1996bx} from the infrared sector of the theory. These 
non-perturbative effects are parameterized by non-vanishing 
local condensates, such as   $D=3$ quark condensate 
\ensuremath{\left\langle \bar{\psi} \psi \right\rangle} and 
 $D=4$ gluon 
condensate \ensuremath{\left\langle G_{\mu\nu}^2\right\rangle}. The quark 
condensate is related to the spontaneous 
breaking of chiral symmetry while the gluon condensate is the manifestation 
of the broken scale invariance.  
Through the operator product expansion (OPE), these condensates appear with 
power corrections in physical observables which provide non-perturbative 
information in addition to perturbatively calculable radiative corrections. 
This strategy has met with success in QCD sum rule calculations both at 
zero~\cite{shifman1992vacuum,* narison2004qcd} and non-zero 
temperature~\cite{Bochkarev:1985ex}. The operator product expansion 
has also been used to investigate nonperturbative corrections in $N$ point Green's functions 
at zero temperature~\cite{Lavelle:1988eg, * Lavelle:1990xg, * Lavelle:1988fj, * 
Bagan:1989dr} as well as  at finite  temperature~\cite{Schaefer:1998wd, 
Chakraborty:2011uw, Schmidt:1999je,Mustafa:1999jz}. 

Of late, there has been much interest on  the existence of a dimension two 
BRST invariant gluon condensate \ensuremath{\left\langle A^2 
\right\rangle}~\cite{Boucaud:2000nd, Gubarev:2000eu, Gubarev:2000nz, 
Kondo:2001nq,Verschelde:2001ia}. This operator is not gauge invariant and does not show up 
in QCD sum rule calculations which deal with gauge invariant correlation 
functions.   However, the situation is different when one considers gauge 
variant quantities such as propagator. In principle, any condensate with 
appropriate quantum numbers may appear in the OPE of Green's 
functions~\cite{Lavelle:1992yh}. One can formally construct a 
gauge invariant, albeit non-local, \ensuremath{\left\langle A^2 \right\rangle} 
by averaging it over all possible gauge transformations. A non-local but 
gauge invariant operator 
does not make much sense  from the perspective of operator product expansion. 
Alternatively one may evaluate the operator  in a specific gauge. In 
particular, it has been shown that the bulk averaged 
\ensuremath{\left\langle A^2 \right\rangle} attains a minimum in Landau gauge. 
Lattice simulations show that 
\ensuremath{\left\langle A^2 \right\rangle} condensate in Landau gauge at zero 
temperature is in fact a large 
quantity~\cite{Boucaud:2000nd,RuizArriola:2004en,Dudal:2010tf}. 

 At finite temperature, the temporal (electric)  and spatial (magnetic) 
 components of the gauge field fluctuations have different size in general.
In fact, the scenario  of dynamical generation of a 
$\langle A_0^2\rangle_T$ condensate at finite temperature has been around 
 in the literature for some time. Possible phenomenological implication of 
such a scenario have been discussed in the context of  static screening of 
chromo-electric and chromo-magnetic fields~\cite{Nadkarni:1986as,Xu:2011ud} (see also~\cite{Rebhan:1994mx} 
and references therein), dilepton production rate~\cite{Lee:1998nz}, Polyakov loop correlator 
~\cite{Megias:2005ve}, trace anomaly~\cite{Megias:2009mp} or physics of heavy 
quarks~\cite{Megias:2007pq,Riek:2010fk} in the quark gluon plasma.  
Nonzero values of electric \ensuremath{\left\langle A_0^2 \right\rangle_T} 
and \ensuremath{\left\langle A_i^2 \right\rangle_T} have been confirmed on 
lattice~\cite{Mandula:1987cp,
Chernodub:2008kf} for \ensuremath{SU_c(2)}. So far no lattice  
measurements for these condensates have been performed for $SU(3)$ but it is 
expected to be qualitatively different.  

With the understanding that $D=2$ gluon condensates have important 
role in the  
nonperturbative dynamics of glue matter, we aim in this  letter  to
 explicitly evaluate their contribution to the in-medium 
quark, gluon and ghost propagation above the deconfinement temperature.

\section[npgpgc]{Non perturbative gluon propagator and gluon condensates}

In QCD, polarization tensor is not 
transverse in general, \(K^\mu \Pi_{\mu \nu}\left(K\right) \neq 0\). 
For an $O(3)$ invariant gauge fixing condition, the most  
general tensorial structure of the in-medium gluon self-energy can be 
written as (we omit trivial color factor)~\cite{Heinz:1986kz}  
\begin{equation}
\label{self_energy_general}
\pi_{\mu\nu} \left(\omega, k\right) = \pi_l\left(\omega, k\right) 
\mathcal{P}^l_{\mu\nu}  +  \pi_t\left(\omega, k\right) \mathcal{P}^t_{\mu\nu} + 
\pi_m \left(\omega, k\right) \mathcal{M}_{\mu\nu} + 
\tilde{\pi} \left(\omega, k\right) l_{\mu\nu}\,. 
\end{equation}
Here, $\omega$ and $k$ are the Lorentz invariant single particle energy and momentum respectively, 
\begin{eqnarray}
\omega &=& u \cdot K\,, \nonumber \\
k &=& \left[\left(u \cdot K\right)^2 - K^2\right]^\frac{1}{2}\,,
\end{eqnarray}
and $u^\mu$ is the four velocity of the heat bath. In the rest frame of the medium $u^\mu = \delta^\mu_0$.  
The projection operators are defined as~\cite{Weldon:1982aq,Heinz:1986kz},
\begin{subequations}
\begin{align}
\mathcal{P}^l_{\mu\nu} &= -\frac{1}{k^2 K^2}\left(k^2 u_\mu +
\omega\widetilde{K_\mu}\right) 
\left(k^2 u_\nu + \omega\widetilde{K_\nu}\right) = \frac{K^2}{\widetilde{K}^2}
 \bar{u}^\mu \bar{u}^\nu \,, \\
\mathcal{P}^t_{\mu\nu} &= \eta_{\mu\nu} - u_\mu u_\nu
- \frac{\widetilde{K_\mu} \widetilde{K_\nu}}{\widetilde{K}^2}\,,\\
\mathcal{M}_{\mu\nu} &= - \frac{1}{\sqrt{-2 \widetilde{K}^2}} \left(
\bar{u}_\mu K_\nu + \bar{u}_\nu K_\mu
\right)\,,\\
l_{\mu\nu} &= \frac{K_\mu K_\nu}{K^2}\,.
\end{align} 
\end{subequations}
where  \ensuremath{\widetilde{K_\mu} = K_\mu - \omega u_\mu} and $\bar{u}^\mu
= u^\mu - \frac{\omega}{K^2}K^\mu$. The projectors $\mathcal{P}_l^{\mu \nu}$
and $\mathcal{P}_t^{\mu \nu}$ are transverse with respect to $K^\mu$ and  
$\mathcal{M}^{\mu \nu}$ satisfies a weaker condition 
$K_\mu \mathcal{M}^{\mu \nu} K_\nu = 0$. The scalar structure functions in the self 
energy are extracted through appropriate projections,   
\begin{subequations}
\begin{align}
\pi_l  &= \mathcal{P}_l^{\mu\nu} \pi_{\mu\nu}\,, \\
\pi_t  &= \frac{1}{2} \mathcal{P}_t^{\mu\nu} \pi_{\mu\nu}\,, \\
\pi_m  &= - \mathcal{M}^{\mu\nu} \pi_{\mu\nu}\,, \\
\widetilde{\pi}  &= l^{\mu\nu} \pi_{\mu\nu}\,.
\end{align} 
\label{pi_structures}
\end{subequations}

Now from (\ref{self_energy_general}), the most general form of the gluon 
propagator $\mathcal{D}_{\mu\nu} = \mathcal{D}_{0,\mu\nu}\left(1 +
\pi_{\mu\nu} \mathcal{D}_{0,\mu\nu}\right)^{-1}$ can be written as,  
\begin{eqnarray}
\mathcal{D}_{\mu\nu} 
&=& - \frac{\mathcal{P}_{\mu\nu}^t}{K^2 - \pi_t} - \frac{2}{2\left(K^2
  - \pi_l\right)\left(\xi^{-1}K^2 - \widetilde{\pi}\right) + 
\pi_m^2} \nonumber \\
& \times & \left[\left(\xi^{-1} K^2 - \widetilde{\pi}\right) 
\mathcal{P}^l_{\mu\nu} + \pi_m \mathcal{M}_{\mu\nu} + \left(K^2 - \pi_l\right)
l_{\mu\nu}\right]\,.
\label{non_tr_prop}
\end{eqnarray}
In covariant gauges, the Slavnov Taylor identity reads
$K^\mu \mathcal{D}_{\mu\nu} K^\nu =  K^\mu \mathcal{D}_{0,\mu\nu} K^\nu =
-\xi$, we get   
\begin{equation}
\label{sti_rel}
\frac{2\left(K^2 -\pi_l\right) K^2}{2\left(K^2 -\pi_l\right)\left(\xi^{-1} K^2 -
\tilde{\pi}\right) + \pi_m^2} = \xi\,.
\end{equation} 
For $\xi \neq 0$. Eq.~(\ref{sti_rel}) can be written as, 
\begin{equation}
\label{sti2}
\left(K^2 - \pi_l\right) \tilde{\pi} = \frac{\pi_m^2}{2}
\end{equation}
This leaves three independent components in (\ref{non_tr_prop}). So the 
most general form of the `non perturbative' gluon propagator should be, 
\begin{eqnarray}
D^{ab}_{\mu\nu}  &\stackrel{\rm def}{=}& \mathcal{D}^{ab,{\rm }}_{\mu\nu}  - 
\mathcal{D}^{ab,{\rm pert}}_{\mu\nu}\nonumber\\
&=& P^l_{\mu\nu}
D_l\left(\omega,k\right) + P^t_{\mu\nu}
D_t\left(\omega,k\right) + \mathcal{M}_{\mu\nu} D_m\left(\omega,k\right)
\,, \label{non-pt-gluon-prop}
\end{eqnarray}
in an obvious notation. Note that, $D_m$ is absent in covariant gauges if 
1) $\xi = 0$ (Landau gauge) or 2) the self energy is transverse $\pi_m =
\tilde{\pi} = 0$.  

\begin{figure}[t]
\begin{minipage}[t]{\textwidth}
\centering{
\includegraphics[width=.50\textwidth,keepaspectratio]{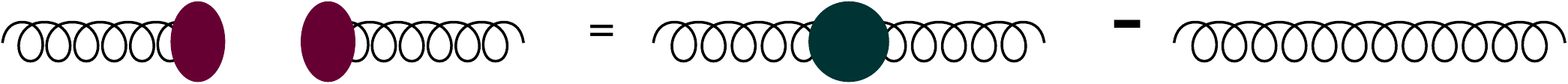}
}
\end{minipage}
\caption{\label{non_pt_gluon_prop} Graphical representation of Eq. 
\ref{non-pt-gluon-prop}. Full and nonperturbative propagators are denoted by circular and oval 
blobs respectively.}
\end{figure}

Condensates of various dimensions can be related to the moments of 
the non-perturbative gluon propagator. 
In the rest frame of the medium (\ensuremath{u^\mu = \delta^\mu_0}), 
the dimension two condensates are given by, 
\begin{subequations}
\label{A2_condensate}
\begin{align}
\left\langle  A_0^2 \right\rangle_T & = 
 - T\left(N_c^2 -1\right)\int\,\frac{d^3k}{\left(2\pi\right)^3}
D_l\left(0, k\right)\,, \\ 
\left\langle  A_i^2 \right\rangle_T & =  2 T\ \left(N_c^2 -1\right)
\int\,\frac{d^3k}{\left(2\pi\right)^3} D_t\left(0, k\right)\,,
\end{align}
\end{subequations}
where $N_c$ is the number of color. We have restricted here to the lowest \
Matsubara mode ($k_0 = 0$) in the spirit of planewave method~\cite{Schaefer:1998wd,Chakraborty:2011uw}. 

\section{Quark self energy}
\begin{figure}[!htbp]
\begin{minipage}[t]{\textwidth}
\centering{
\includegraphics[width=.25\textwidth,keepaspectratio]{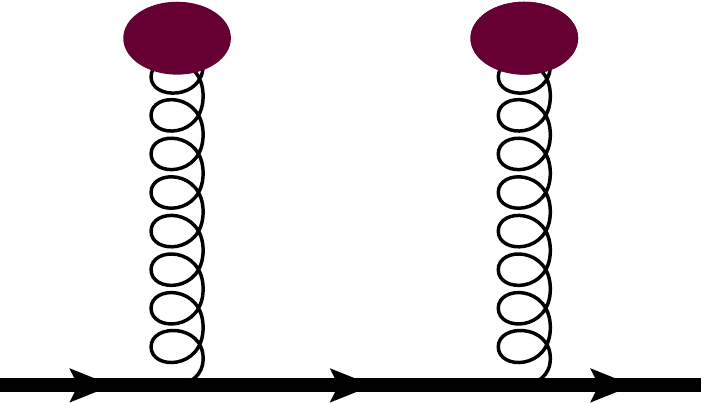}
}
\end{minipage}
\caption{\label{quark_self_gluon_condensate} Gluon condensate contribution 
to quark self energy.}
\end{figure}
The $D=4$ gluon condensate contribution to the quark self energy has been 
studied in~\cite{Schaefer:1998wd}. Here we shall evaluate the $D=2$ condensate 
contribution.  
The general expression for in-medium fermion self energy in the chiral limit 
is given by, 
\begin{equation}
\Sigma \left(P\right) = - a\left(\omega,p\right) \slashed{P} - b\left(\omega,p\right)
\slashed{u}\,,
\end{equation}
where the scalar functions are given by, 
\begin{subequations}
\begin{align}
a \left(\omega,p\right) &= \frac{1}{4p^2}\left[\mathrm{Tr}\left(\slashed{P}\Sigma\right)
- \omega \mathrm{Tr}\left(\slashed{u}\Sigma\right)\right]\,,\\
b \left(\omega,p\right) &= \frac{1}{4p^2}\left[P^2\mathrm{Tr}\left(\slashed{u}\Sigma\right)
- \omega \mathrm{Tr}\left(\slashed{P}\Sigma\right)\right]\,.
\end{align}
\end{subequations}
Using the plane wave method~\cite{Reinders:1984sr,Schaefer:1998wd} and 
nonperturbative propagator from Eq.~(\ref{non-pt-gluon-prop}), we get from 
Fig.~\ref{quark_self_gluon_condensate} 
\begin{subequations}
\begin{align}
a\left(p_0,p\right) &= -\frac{2\pi\alpha_s}{N_c P^2} \left[\frac{1}{3}
\left\langle A_i^2 \right\rangle_T - \left\langle A_0^2 \right\rangle_T
\right]\,, \\
b\left(p_0,p\right) &= -\frac{2\pi\alpha_s p_0}{N_c P^2} \left[
2\left\langle A_0^2 \right\rangle_T +\frac{2}{3} \left\langle A_i^2 \right\rangle_T
\right]\,. 
\end{align}
\end{subequations}
Note that, $a$ and $b$ contain nonperturbative contribution from 
 $D=4$ gluon condensate and other higher dimensional condensates 
and perturbative contribution given by HTL corrections as,  
\begin{eqnarray}
a &=& a^{\left\langle A^2 \right\rangle} +  a^{\left\langle G^2 \right\rangle} + \cdots + 
 a^{\rm HTL} \,,\nonumber\\
b &=& b^{\left\langle A^2 \right\rangle} +  b^{\left\langle G^2 \right\rangle} + \cdots + 
 b^{\rm HTL }\,.
\end{eqnarray}
Then, the chiral quark propagator \ensuremath{S^{-1}(P) = \slashed{P} - \Sigma}
follows as
\begin{eqnarray}
S\left(p_0,p\right) = \frac{\gamma_0 -\gamma\cdot\hat{\vec{p}}}{2D_+\left(p_0,p\right)} + \frac{\gamma_0 + \gamma\cdot\hat{\vec{p}}}{2D_-\left(p_0,p\right)}\,,
\end{eqnarray}
where,
\begin{equation}
D_{\pm}\left(p_0,p\right) = \left(-p_0 \pm p \right)\left(1 + a\right) - b\,.
\end{equation} 

\section{Gluon self Energy}
\begin{figure}[!h]
\begin{minipage}[t]{\textwidth}
\hfill
\begin{minipage}[t]{.45\textwidth}
\begin{flushright}
\includegraphics[width=.5\columnwidth]{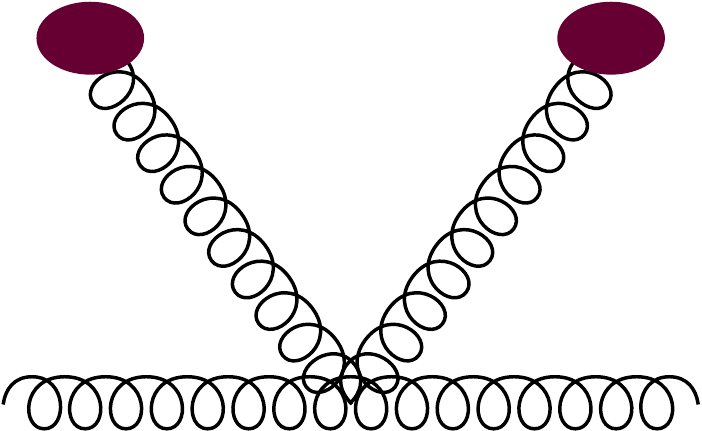}
\end{flushright}
\end{minipage}
\hfill
\begin{minipage}[t]{.45\textwidth}
\begin{flushleft}
\includegraphics[width=.5\columnwidth]{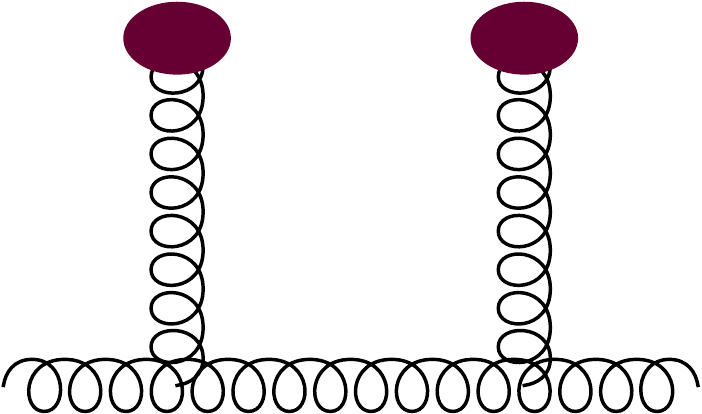}
\end{flushleft}
\end{minipage}
\end{minipage}
\caption{\label{3g_and_4g} Gluon self energy with non-perturbative gluon 
propagator. }
\end{figure}
The gluon self energy with $D=2$ gluon condensate follows from 
Fig.~\ref{3g_and_4g}. Various scalar structure functions in the self energy 
as given in~(\ref{pi_structures}) are obtained as,  
\begin{eqnarray}
\pi_l^{\left\langle A^2 \right\rangle}  &=& 
f_i  \frac{\alpha_s}{\pi}
\left\langle A_i^2 \right \rangle_T - 
f_0  \frac{\alpha_s}{\pi}\left\langle A_0^2 \right \rangle_T\,, \\
\pi_t^{\left\langle A^2 \right\rangle} &=& 
 g_i  \frac{\alpha_s}{\pi}
\left\langle A_i^2 \right \rangle_T - 
 g_0 \frac{\alpha_s}{\pi} \left\langle A_0^2 \right\rangle_T\,, \\
\pi_m^{\left\langle A^2 \right\rangle} &=& 
\sqrt{2} h_i  \frac{\alpha_s}{\pi}\left\langle A_i^2 \right \rangle_T - 
\sqrt{2} h_0  \frac{\alpha_s}{\pi}\left\langle A_0^2 \right
\rangle_T\,,\\
\widetilde{\pi}^{\left\langle A^2 \right\rangle} &=& 
w_i  \frac{\alpha_s}{\pi}\left\langle A_i^2 \right \rangle_T - 
w_0  \frac{\alpha_s}{\pi}\left\langle A_0^2 \right
\rangle_T\,. 
\end{eqnarray}
with 
\begin{subequations}
\begin{align}
f_0 \left(p_0,p\right) & =  \frac{4 \pi^2 N_c}{\left(N_c^2 -1\right)} \left\{
\left(-\frac{\vec{p}^2}{P^2} + \frac{4p_0^2}{P^2} + \frac{p_0^2\vec{p}^2}{P^4}
- \frac{4p_0^4}{P^4} -  \frac{3p_0^4}{\vec{p}^2 P^2} + \frac{3p_0^6}{\vec{p}^2
  P^4}\right) \right.\nonumber \\ 
& \left. + \chi \left( \frac{\vec{p}^2}{P^2} - \frac{2 p_0^2}{P^2} -
 \frac{2 p_0^2 \vec{p}^2}{P^4} + \frac{4p_0^4}{P^4} +  \frac{p_0^4}{\vec{p}^2
   P^2}  + \frac{p_0^4 \vec{p}^2}{P^6} + \frac{2p_0^6}{P^6} 
\right. \right. \nonumber \\
& \left. \left. - \frac{2p_0^6}{\vec{p}^2 P^4} + \frac{p_0^8}{\vec{p}^2 P^6}
\right) \right\}\,, \\
f_i \left(p_0,p\right) & =  \frac{2 \pi^2 N_c}{\left(N_c^2 -1\right)} \left\{
\left(-\frac{2\vec{p}^2}{P^2} + \frac{2p_0^2}{P^2} - \frac{4 \vec{p}^4}{3 P^4}
+ \frac{4 p_0^2 \vec{p}^2}{P^4} -  \frac{8 p_0^4}{3 P^4} \right) + \chi \left(
 \frac{p_0^2 \vec{p}^2}{P^4} + \frac{p_0^2 \vec{p}^4}{3 P^6}\right. \right. \nonumber \\ 
& \left. \left. - \frac{2 p_0^4}{P^4} -
 \frac{5 p_0^4 \vec{p}^2}{3 P^6} + \frac{7 p_0^6}{3 P^6} +  \frac{p_0^6}{\vec{p}^2
   P^4}  - \frac{p_0^8} {\vec{p}^2 P^6} 
\right) \right\}\,,
\end{align}
\end{subequations}

\begin{subequations}
\begin{align}
g_0 \left(p_0,p\right) & =  \frac{4 \pi^2 N_c}{\left(N_c^2 -1\right)} \left\{
\left(-1 + \frac{11 p_0^2}{2 P^2}\right) + \chi \left(-\frac{p_0^2}{P^2} + 
\frac{p_0^4}{2 P^4} - \frac{p_0^4}{2 \vec{p}^2 P^2} - 
\frac{p_0^6}{2 \vec{p}^2 P^4}
\right)\right\}\,\\
g_i \left(p_0,p\right) & =  \frac{2 \pi^2 N_c}{\left(N_c^2 -1\right)} \left\{
\left(\frac{5}{2} + \frac{\vec{p}^2}{P^2} - \frac{11 p_0^2}{3 P^2}\right) + 
\chi\left(\frac{\vec{p}^2}{6 P^2} + 
\frac{p_0^2 \vec{p}^2}{6 P^4} + \frac{2p_0^4}{3 P^4} - 
\frac{p_0^4}{2 \vec{p}^2 P^4} + \frac{p_0^6}{2 \vec{p}^2 P^4}
\right)\right\}\,
\end{align}
\end{subequations}

\begin{subequations}
\begin{align}
h_0 \left(p_0,p\right) & =  \frac{4 \pi^2 N_c}{\left(N_c^2 -1\right)} \left\{
-\frac{p_0}{\left|\vec{p}\right|} +\frac{p_0^3}{\left|\vec{p}P^2\right|}
\right\}\,,\\
h_i \left(p_0,p\right) & =  \frac{2 \pi^2 N_c}{\left(N_c^2 -1\right)} \left\{
\frac{p_0}{\left|\vec{p}\right|} + \frac{p_0 \left|\vec{p}\right|}{3P^2} -
\frac{p_0^3}{\left|\vec{p}P^2\right|}\right\}\,,
\end{align}
\end{subequations}

\begin{subequations}
\begin{align}
w_0 \left(p_0,p\right) & =  0\,,\\
w_i \left(p_0,p\right) & =  0.
\end{align}
\end{subequations}
Here $\chi = 1 - \xi$ and $\xi$ is gauge parameter.  

At zero temperature, $k^\mu \pi^{\langle A^2 \rangle}_{\mu\nu} = 0 $. We find
that the transversality is weaker at finite temperature 
$k^\mu \pi^{\langle A^2 \rangle}_{\mu\nu}k^\nu = 0$. 
The $D=4$ condensate contribution to the gluon self energy will be 
presented elsewhere~\cite{cmt:2011b}.   


\section{Ghost self energy}
\begin{figure}[!htbp]
\begin{minipage}[t]{\textwidth}
\centering{
\includegraphics[width=.25\textwidth,keepaspectratio]{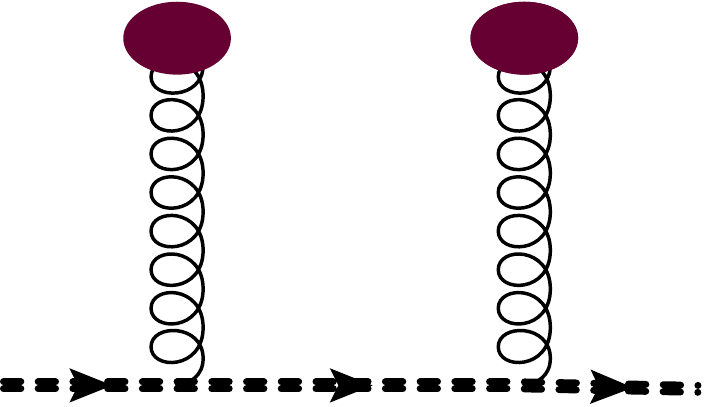}
}
\end{minipage}
\caption{\label{ghost_self_gluon_condensate}Gluon condensate contribution to ghost self energy}
\end{figure}
Similarly, $D=2$ gluon condensate contribution to ghost self energy in 
covariant gauge can be evaluated from Fig.~\ref{ghost_self_gluon_condensate} as,
\begin{equation}
\Pi \left(p_0, p\right) = \frac{4 \pi \alpha_s N_c}{N_c^2 - 1} \left[
\frac{p_0^2}{P^2} \left\langle A_0^2 \right\rangle_T - \frac{1}{2} \left(
\frac{4}{3} - \frac{p_0^2}{P^2}\right) \left\langle A_i^2 \right\rangle_T\right]\,.
\end{equation}
There is no HTL correction for ghost, but ghost can receive non-perturbative 
mass  corrections in a background characterized by non-vanishing 
condensates. One can similarly calculate Wilson coefficients of 
$D=4$ gluon condensates to ghost self energy. 
 To fix 
these coefficients uniquely, one has to go beyond one loop and calculate 
Wilson coefficients of $D=4$ ghost-anti ghost 
$\left(\bar{\eta} \Box \eta\right)$, and mixed ghost-gluon 
condensate $\left(f^{abc} \partial_\mu \bar{\eta}^a A^{\mu,b} \eta^c\right)$. 
This is similar to gluon self energy calculation at one 
loop~\cite{Chakraborty:2011uw} and details will be presented 
elsewhere~\cite{cmt:2011b}.    

\section{Summary}
We have calculated $D=2$ gluon condensate contribution to quark, gluon and 
ghost self 
energies at finite temperature. With the values of condensates 
taken as input from lattice QCD, one can quantitatively predict 
 the analytical structure of these propagators near the transition 
 temperature. As correlated applications, this will be a good starting point 
to estimate nonperturbative dilepton rate, transport properties of both 
heavy and light quarks etc. in the deconfined system of quark-gluon matter. 
 
\bibliography{dim2refs}{}

\begin{thebibliography}{36}%
\makeatletter
\providecommand \@ifxundefined [1]{%
 \@ifx{#1\undefined}
}%
\providecommand \@ifnum [1]{%
 \ifnum #1\expandafter \@firstoftwo
 \else \expandafter \@secondoftwo
 \fi
}%
\providecommand \@ifx [1]{%
 \ifx #1\expandafter \@firstoftwo
 \else \expandafter \@secondoftwo
 \fi
}%
\providecommand \natexlab [1]{#1}%
\providecommand \enquote  [1]{``#1''}%
\providecommand \bibnamefont  [1]{#1}%
\providecommand \bibfnamefont [1]{#1}%
\providecommand \citenamefont [1]{#1}%
\providecommand \href@noop [0]{\@secondoftwo}%
\providecommand \href [0]{\begingroup \@sanitize@url \@href}%
\providecommand \@href[1]{\@@startlink{#1}\@@href}%
\providecommand \@@href[1]{\endgroup#1\@@endlink}%
\providecommand \@sanitize@url [0]{\catcode `\\12\catcode `\$12\catcode
  `\&12\catcode `\#12\catcode `\^12\catcode `\_12\catcode `\%12\relax}%
\providecommand \@@startlink[1]{}%
\providecommand \@@endlink[0]{}%
\providecommand \url  [0]{\begingroup\@sanitize@url \@url }%
\providecommand \@url [1]{\endgroup\@href {#1}{\urlprefix }}%
\providecommand \urlprefix  [0]{URL }%
\providecommand \Eprint [0]{\href }%
\@ifxundefined \urlstyle {%
  \providecommand \doi  [0]{\begingroup \@sanitize@url \@doi}%
  \providecommand \@doi [1]{\endgroup \@@startlink {\doibase
  #1}doi:\discretionary {}{}{}#1\@@endlink }%
}{%
  \providecommand \doi  [0]{doi:\discretionary{}{}{}\begingroup
  \urlstyle{rm}\Url }%
}%
\providecommand \doibase [0]{http://dx.doi.org/}%
\providecommand \Doi [0]{\begingroup \@sanitize@url \@Doi }%
\providecommand \@Doi  [1]{\endgroup\@@startlink{\doibase#1}\@@Doi}%
\providecommand \@@Doi [1]{#1\@@endlink}%
\providecommand \selectlanguage [0]{\@gobble}%
\providecommand \bibinfo  [0]{\@secondoftwo}%
\providecommand \bibfield  [0]{\@secondoftwo}%
\providecommand \translation [1]{[#1]}%
\providecommand \BibitemOpen [0]{}%
\providecommand \bibitemStop [0]{}%
\providecommand \bibitemNoStop [0]{.\EOS\space}%
\providecommand \EOS [0]{\spacefactor3000\relax}%
\providecommand \BibitemShut  [1]{\csname bibitem#1\endcsname}%
\bibitem [{\citenamefont {Aoki}\ \emph {et~al.}(2006)\citenamefont {Aoki},
  \citenamefont {Endrodi}, \citenamefont {Fodor}, \citenamefont {Katz},\ and\
  \citenamefont {Szabo}}]{Aoki:2006we}%
  \BibitemOpen
  \bibfield  {author} {\bibinfo {author} {\bibfnamefont {Y.}~\bibnamefont
  {Aoki}}, \bibinfo {author} {\bibfnamefont {G.}~\bibnamefont {Endrodi}},
  \bibinfo {author} {\bibfnamefont {Z.}~\bibnamefont {Fodor}}, \bibinfo
  {author} {\bibfnamefont {S.}~\bibnamefont {Katz}}, \ and\ \bibinfo {author}
  {\bibfnamefont {K.}~\bibnamefont {Szabo}},\ }\Doi {10.1038/nature05120}
  {\bibfield  {journal} {\bibinfo  {journal} {Nature},\ }\textbf {\bibinfo
  {volume} {443}},\ \bibinfo {pages} {675} (\bibinfo {year}
  {2006})}\BibitemShut {NoStop}%
\bibitem [{\citenamefont {Kapusta}\ and\ \citenamefont
  {Gale}(2006)}]{kapusta2006finite}%
  \BibitemOpen
  \bibfield  {author} {\bibinfo {author} {\bibfnamefont {J.}~\bibnamefont
  {Kapusta}}\ and\ \bibinfo {author} {\bibfnamefont {C.}~\bibnamefont {Gale}},\
  }\href@noop {} {\emph {\bibinfo {title} {Finite-temperature field theory:
  Principles and applications}}}\ (\bibinfo  {publisher} {Cambridge Univ Pr},\
  \bibinfo {year} {2006})\BibitemShut {NoStop}%
\bibitem [{\citenamefont {Boyd}\ \emph {et~al.}(1996)\citenamefont {Boyd},
  \citenamefont {Engels}, \citenamefont {Karsch}, \citenamefont {Laermann},
  \citenamefont {Legeland} \emph {et~al.}}]{Boyd:1996bx}%
  \BibitemOpen
  \bibfield  {author} {\bibinfo {author} {\bibfnamefont {G.}~\bibnamefont
  {Boyd}}, \bibinfo {author} {\bibfnamefont {J.}~\bibnamefont {Engels}},
  \bibinfo {author} {\bibfnamefont {F.}~\bibnamefont {Karsch}}, \bibinfo
  {author} {\bibfnamefont {E.}~\bibnamefont {Laermann}}, \bibinfo {author}
  {\bibfnamefont {C.}~\bibnamefont {Legeland}},  \emph {et~al.},\ }\Doi
  {10.1016/0550-3213(96)00170-8} {\bibfield  {journal} {\bibinfo  {journal}
  {Nucl. Phys.},\ }\textbf {\bibinfo {volume} {B469}},\ \bibinfo {pages} {419}
  (\bibinfo {year} {1996})}\BibitemShut {NoStop}%
\bibitem [{\citenamefont {Shifman}(1992)}]{shifman1992vacuum}%
  \BibitemOpen
  \bibfield  {author} {\bibinfo {author} {\bibfnamefont {M.}~\bibnamefont
  {Shifman}},\ }\href@noop {} {\emph {\bibinfo {title} {{Vacuum structure and
  QCD sum rules}}}}\ (\bibinfo  {publisher} {North-Holland},\ \bibinfo {year}
  {1992})\BibitemShut {NoStop}%
\bibitem [{\citenamefont {Narison}(2004)}]{narison2004qcd}%
  \BibitemOpen
  \bibfield  {author} {\bibinfo {author} {\bibfnamefont {S.}~\bibnamefont
  {Narison}},\ }\href@noop {} {\emph {\bibinfo {title} {{QCD as a theory of
  hadrons: from partons to confinement}}}}\ (\bibinfo  {publisher}
  {Cambridge},\ \bibinfo {year} {2004})\BibitemShut {NoStop}%
\bibitem [{\citenamefont {Bochkarev}\ and\ \citenamefont
  {Shaposhnikov}(1986)}]{Bochkarev:1985ex}%
  \BibitemOpen
  \bibfield  {author} {\bibinfo {author} {\bibfnamefont {A.}~\bibnamefont
  {Bochkarev}}\ and\ \bibinfo {author} {\bibfnamefont {M.}~\bibnamefont
  {Shaposhnikov}},\ }\Doi {10.1016/0550-3213(86)90209-9} {\bibfield  {journal}
  {\bibinfo  {journal} {Nucl.\ Phys.},\ }\textbf {\bibinfo {volume} {B268}},\
  \bibinfo {pages} {220} (\bibinfo {year} {1986})}\BibitemShut {NoStop}%
\bibitem [{\citenamefont {Lavelle}\ and\ \citenamefont
  {Schaden}(1988)}]{Lavelle:1988eg}%
  \BibitemOpen
  \bibfield  {author} {\bibinfo {author} {\bibfnamefont {M.~J.}\ \bibnamefont
  {Lavelle}}\ and\ \bibinfo {author} {\bibfnamefont {M.}~\bibnamefont
  {Schaden}},\ }\Doi {10.1016/0370-2693(88)90433-9} {\bibfield  {journal}
  {\bibinfo  {journal} {Phys. Lett.},\ }\textbf {\bibinfo {volume} {B208}},\
  \bibinfo {pages} {297} (\bibinfo {year} {1988})}\BibitemShut {NoStop}%
\bibitem [{\citenamefont {Lavelle}\ and\ \citenamefont
  {Schaden}(1990)}]{Lavelle:1990xg}%
  \BibitemOpen
  \bibfield  {author} {\bibinfo {author} {\bibfnamefont {M.}~\bibnamefont
  {Lavelle}}\ and\ \bibinfo {author} {\bibfnamefont {M.}~\bibnamefont
  {Schaden}},\ }\Doi {10.1016/0370-2693(90)90635-J} {\bibfield  {journal}
  {\bibinfo  {journal} {Phys. Lett.},\ }\textbf {\bibinfo {volume} {B246}},\
  \bibinfo {pages} {487} (\bibinfo {year} {1990})}\BibitemShut {NoStop}%
\bibitem [{\citenamefont {Lavelle}\ and\ \citenamefont
  {Schaden}(1989)}]{Lavelle:1988fj}%
  \BibitemOpen
  \bibfield  {author} {\bibinfo {author} {\bibfnamefont {M.}~\bibnamefont
  {Lavelle}}\ and\ \bibinfo {author} {\bibfnamefont {M.}~\bibnamefont
  {Schaden}},\ }\Doi {10.1016/0370-2693(89)90095-6} {\bibfield  {journal}
  {\bibinfo  {journal} {Phys. Lett.},\ }\textbf {\bibinfo {volume} {B217}},\
  \bibinfo {pages} {551} (\bibinfo {year} {1989})}\BibitemShut {NoStop}%
\bibitem [{\citenamefont {Bagan}\ and\ \citenamefont
  {Steele}(1989)}]{Bagan:1989dr}%
  \BibitemOpen
  \bibfield  {author} {\bibinfo {author} {\bibfnamefont {E.}~\bibnamefont
  {Bagan}}\ and\ \bibinfo {author} {\bibfnamefont {T.~G.}\ \bibnamefont
  {Steele}},\ }\Doi {10.1016/0370-2693(89)90303-1} {\bibfield  {journal}
  {\bibinfo  {journal} {Phys. Lett.},\ }\textbf {\bibinfo {volume} {B226}},\
  \bibinfo {pages} {142} (\bibinfo {year} {1989})}\BibitemShut {NoStop}%
\bibitem [{\citenamefont {Schaefer}\ and\ \citenamefont
  {Thoma}(1999)}]{Schaefer:1998wd}%
  \BibitemOpen
  \bibfield  {author} {\bibinfo {author} {\bibfnamefont {A.}~\bibnamefont
  {Schaefer}}\ and\ \bibinfo {author} {\bibfnamefont {M.~H.}\ \bibnamefont
  {Thoma}},\ }\Doi {10.1016/S0370-2693(99)00186-0} {\bibfield  {journal}
  {\bibinfo  {journal} {Phys.\ Lett.},\ }\textbf {\bibinfo {volume} {B451}},\
  \bibinfo {pages} {195} (\bibinfo {year} {1999})}\BibitemShut {NoStop}%
\bibitem [{\citenamefont {Chakraborty}\ \emph {et~al.}(2012)\citenamefont
  {Chakraborty}, \citenamefont {Mustafa},\ and\ \citenamefont
  {Thoma}}]{Chakraborty:2011uw}%
  \BibitemOpen
  \bibfield  {author} {\bibinfo {author} {\bibfnamefont {P.}~\bibnamefont
  {Chakraborty}}, \bibinfo {author} {\bibfnamefont {M.~G.}\ \bibnamefont
  {Mustafa}}, \ and\ \bibinfo {author} {\bibfnamefont {M.~H.}\ \bibnamefont
  {Thoma}},\ }\Doi {10.1103/PhysRevD.85.056002} {\bibfield  {journal} {\bibinfo
   {journal} {Phys.\ Rev.},\ }\textbf {\bibinfo {volume} {D85}},\ \bibinfo
  {pages} {056002} (\bibinfo {year} {2012})}\BibitemShut {NoStop}%
\bibitem [{\citenamefont {Schmidt}\ and\ \citenamefont
  {Yang}(1999)}]{Schmidt:1999je}%
  \BibitemOpen
  \bibfield  {author} {\bibinfo {author} {\bibfnamefont {I.}~\bibnamefont
  {Schmidt}}\ and\ \bibinfo {author} {\bibfnamefont {J.-J.}\ \bibnamefont
  {Yang}},\ }\Doi {10.1016/S0370-2693(99)01201-0} {\bibfield  {journal}
  {\bibinfo  {journal} {Phys.\ Lett.},\ }\textbf {\bibinfo {volume} {B468}},\
  \bibinfo {pages} {138} (\bibinfo {year} {1999})}\BibitemShut {NoStop}%
\bibitem [{\citenamefont {Mustafa}\ \emph {et~al.}(2000)\citenamefont
  {Mustafa}, \citenamefont {Schaefer},\ and\ \citenamefont
  {Thoma}}]{Mustafa:1999jz}%
  \BibitemOpen
  \bibfield  {author} {\bibinfo {author} {\bibfnamefont {M.~G.}\ \bibnamefont
  {Mustafa}}, \bibinfo {author} {\bibfnamefont {A.}~\bibnamefont {Schaefer}}, \
  and\ \bibinfo {author} {\bibfnamefont {M.~H.}\ \bibnamefont {Thoma}},\ }\Doi
  {10.1016/S0370-2693(99)01441-0} {\bibfield  {journal} {\bibinfo  {journal}
  {Phys.\ Lett.},\ }\textbf {\bibinfo {volume} {B472}},\ \bibinfo {pages} {402}
  (\bibinfo {year} {2000})}\BibitemShut {NoStop}%
\bibitem [{\citenamefont {Boucaud}\ \emph {et~al.}(2000)\citenamefont
  {Boucaud}, \citenamefont {Le~Yaouanc}, \citenamefont {Leroy}, \citenamefont
  {Micheli}, \citenamefont {Pene} \emph {et~al.}}]{Boucaud:2000nd}%
  \BibitemOpen
  \bibfield  {author} {\bibinfo {author} {\bibfnamefont {P.}~\bibnamefont
  {Boucaud}}, \bibinfo {author} {\bibfnamefont {A.}~\bibnamefont {Le~Yaouanc}},
  \bibinfo {author} {\bibfnamefont {J.}~\bibnamefont {Leroy}}, \bibinfo
  {author} {\bibfnamefont {J.}~\bibnamefont {Micheli}}, \bibinfo {author}
  {\bibfnamefont {O.}~\bibnamefont {Pene}},  \emph {et~al.},\ }\Doi
  {10.1016/S0370-2693(00)01149-7} {\bibfield  {journal} {\bibinfo  {journal}
  {Phys. Lett.},\ }\textbf {\bibinfo {volume} {B493}},\ \bibinfo {pages} {315}
  (\bibinfo {year} {2000})}\BibitemShut {NoStop}%
\bibitem [{\citenamefont {Gubarev}\ \emph {et~al.}(2001)\citenamefont
  {Gubarev}, \citenamefont {Stodolsky},\ and\ \citenamefont
  {Zakharov}}]{Gubarev:2000eu}%
  \BibitemOpen
  \bibfield  {author} {\bibinfo {author} {\bibfnamefont {F.}~\bibnamefont
  {Gubarev}}, \bibinfo {author} {\bibfnamefont {L.}~\bibnamefont {Stodolsky}},
  \ and\ \bibinfo {author} {\bibfnamefont {V.~I.}\ \bibnamefont {Zakharov}},\
  }\Doi {10.1103/PhysRevLett.86.2220} {\bibfield  {journal} {\bibinfo
  {journal} {Phys. Rev. Lett.},\ }\textbf {\bibinfo {volume} {86}},\ \bibinfo
  {pages} {2220} (\bibinfo {year} {2001})}\BibitemShut {NoStop}%
\bibitem [{\citenamefont {Gubarev}\ and\ \citenamefont
  {Zakharov}(2001)}]{Gubarev:2000nz}%
  \BibitemOpen
  \bibfield  {author} {\bibinfo {author} {\bibfnamefont {F.}~\bibnamefont
  {Gubarev}}\ and\ \bibinfo {author} {\bibfnamefont {V.~I.}\ \bibnamefont
  {Zakharov}},\ }\Doi {10.1016/S0370-2693(01)00085-5} {\bibfield  {journal}
  {\bibinfo  {journal} {Phys. Lett.},\ }\textbf {\bibinfo {volume} {B501}},\
  \bibinfo {pages} {28} (\bibinfo {year} {2001})}\BibitemShut {NoStop}%
\bibitem [{\citenamefont {Kondo}(2001)}]{Kondo:2001nq}%
  \BibitemOpen
  \bibfield  {author} {\bibinfo {author} {\bibfnamefont {K.-I.}\ \bibnamefont
  {Kondo}},\ }\Doi {10.1016/S0370-2693(01)00817-6} {\bibfield  {journal}
  {\bibinfo  {journal} {Phys. Lett.},\ }\textbf {\bibinfo {volume} {B514}},\
  \bibinfo {pages} {335} (\bibinfo {year} {2001})}\BibitemShut {NoStop}%
\bibitem [{\citenamefont {Verschelde}\ \emph {et~al.}(2001)\citenamefont
  {Verschelde}, \citenamefont {Knecht}, \citenamefont {Van~Acoleyen},\ and\
  \citenamefont {Vanderkelen}}]{Verschelde:2001ia}%
  \BibitemOpen
  \bibfield  {author} {\bibinfo {author} {\bibfnamefont {H.}~\bibnamefont
  {Verschelde}}, \bibinfo {author} {\bibfnamefont {K.}~\bibnamefont {Knecht}},
  \bibinfo {author} {\bibfnamefont {K.}~\bibnamefont {Van~Acoleyen}}, \ and\
  \bibinfo {author} {\bibfnamefont {M.}~\bibnamefont {Vanderkelen}},\ }\Doi
  {10.1016/S0370-2693(01)00929-7} {\bibfield  {journal} {\bibinfo  {journal}
  {Phys.Lett.},\ }\textbf {\bibinfo {volume} {B516}},\ \bibinfo {pages} {307}
  (\bibinfo {year} {2001})}\BibitemShut {NoStop}%
\bibitem [{\citenamefont {Lavelle}\ and\ \citenamefont
  {Oleszczuk}(1992)}]{Lavelle:1992yh}%
  \BibitemOpen
  \bibfield  {author} {\bibinfo {author} {\bibfnamefont {M.}~\bibnamefont
  {Lavelle}}\ and\ \bibinfo {author} {\bibfnamefont {M.}~\bibnamefont
  {Oleszczuk}},\ }\Doi {10.1142/S0217732392003049} {\bibfield  {journal}
  {\bibinfo  {journal} {Mod. Phys. Lett.},\ }\textbf {\bibinfo {volume} {A7}},\
  \bibinfo {pages} {3617} (\bibinfo {year} {1992})}\BibitemShut {NoStop}%
\bibitem [{\citenamefont {Ruiz~Arriola}\ \emph {et~al.}(2004)\citenamefont
  {Ruiz~Arriola}, \citenamefont {Bowman},\ and\ \citenamefont
  {Broniowski}}]{RuizArriola:2004en}%
  \BibitemOpen
  \bibfield  {author} {\bibinfo {author} {\bibfnamefont {E.}~\bibnamefont
  {Ruiz~Arriola}}, \bibinfo {author} {\bibfnamefont {P.~O.}\ \bibnamefont
  {Bowman}}, \ and\ \bibinfo {author} {\bibfnamefont {W.}~\bibnamefont
  {Broniowski}},\ }\Doi {10.1103/PhysRevD.70.097505} {\bibfield  {journal}
  {\bibinfo  {journal} {Phys.Rev.},\ }\textbf {\bibinfo {volume} {D70}},\
  \bibinfo {pages} {097505} (\bibinfo {year} {2004})}\BibitemShut {NoStop}%
\bibitem [{\citenamefont {Dudal}\ \emph {et~al.}(2010)\citenamefont {Dudal},
  \citenamefont {Oliveira},\ and\ \citenamefont {Vandersickel}}]{Dudal:2010tf}%
  \BibitemOpen
  \bibfield  {author} {\bibinfo {author} {\bibfnamefont {D.}~\bibnamefont
  {Dudal}}, \bibinfo {author} {\bibfnamefont {O.}~\bibnamefont {Oliveira}}, \
  and\ \bibinfo {author} {\bibfnamefont {N.}~\bibnamefont {Vandersickel}},\
  }\Doi {10.1103/PhysRevD.81.074505} {\bibfield  {journal} {\bibinfo  {journal}
  {Phys.Rev.},\ }\textbf {\bibinfo {volume} {D81}},\ \bibinfo {pages} {074505}
  (\bibinfo {year} {2010})},\ \Eprint {http://arxiv.org/abs/1002.2374}
  {1002.2374} \BibitemShut {NoStop}%
\bibitem [{\citenamefont {Nadkarni}(1986)}]{Nadkarni:1986as}%
  \BibitemOpen
  \bibfield  {author} {\bibinfo {author} {\bibfnamefont {S.}~\bibnamefont
  {Nadkarni}},\ }\Doi {10.1103/PhysRevD.34.3904} {\bibfield  {journal}
  {\bibinfo  {journal} {Phys.Rev.},\ }\textbf {\bibinfo {volume} {D34}},\
  \bibinfo {pages} {3904} (\bibinfo {year} {1986})}\BibitemShut {NoStop}%
\bibitem [{\citenamefont {Xu}\ and\ \citenamefont {Huang}(2011)}]{Xu:2011ud}%
  \BibitemOpen
  \bibfield  {author} {\bibinfo {author} {\bibfnamefont {F.}~\bibnamefont
  {Xu}}\ and\ \bibinfo {author} {\bibfnamefont {M.}~\bibnamefont {Huang}},\
  }\href@noop {} { (\bibinfo {year} {2011})},\ \Eprint
  {http://arxiv.org/abs/1111.5152} {arXiv:1111.5152 [hep-ph]} \BibitemShut
  {NoStop}%
\bibitem [{\citenamefont {Rebhan}(1994)}]{Rebhan:1994mx}%
  \BibitemOpen
  \bibfield  {author} {\bibinfo {author} {\bibfnamefont {A.~K.}\ \bibnamefont
  {Rebhan}},\ }\Doi {10.1016/0550-3213(94)90253-4} {\bibfield  {journal}
  {\bibinfo  {journal} {Nucl. Phys.},\ }\textbf {\bibinfo {volume} {B430}},\
  \bibinfo {pages} {319} (\bibinfo {year} {1994})}\BibitemShut {NoStop}%
\bibitem [{\citenamefont {Lee}\ \emph {et~al.}(1999)\citenamefont {Lee},
  \citenamefont {Wirstam}, \citenamefont {Zahed},\ and\ \citenamefont
  {Hansson}}]{Lee:1998nz}%
  \BibitemOpen
  \bibfield  {author} {\bibinfo {author} {\bibfnamefont {C.~H.}\ \bibnamefont
  {Lee}}, \bibinfo {author} {\bibfnamefont {J.}~\bibnamefont {Wirstam}},
  \bibinfo {author} {\bibfnamefont {I.}~\bibnamefont {Zahed}}, \ and\ \bibinfo
  {author} {\bibfnamefont {T.~H.}\ \bibnamefont {Hansson}},\ }\href@noop {}
  {\bibfield  {journal} {\bibinfo  {journal} {Phys. Lett.},\ }\textbf {\bibinfo
  {volume} {B448}},\ \bibinfo {pages} {168} (\bibinfo {year}
  {1999})}\BibitemShut {NoStop}%
\bibitem [{\citenamefont {Megias}\ \emph {et~al.}(2006)\citenamefont {Megias},
  \citenamefont {Ruiz~Arriola},\ and\ \citenamefont {Salcedo}}]{Megias:2005ve}%
  \BibitemOpen
  \bibfield  {author} {\bibinfo {author} {\bibfnamefont {E.}~\bibnamefont
  {Megias}}, \bibinfo {author} {\bibfnamefont {E.}~\bibnamefont
  {Ruiz~Arriola}}, \ and\ \bibinfo {author} {\bibfnamefont {L.}~\bibnamefont
  {Salcedo}},\ }\Doi {10.1088/1126-6708/2006/01/073} {\bibfield  {journal}
  {\bibinfo  {journal} {JHEP},\ }\textbf {\bibinfo {volume} {0601}},\ \bibinfo
  {pages} {073} (\bibinfo {year} {2006})}\BibitemShut {NoStop}%
\bibitem [{\citenamefont {Megias}\ \emph {et~al.}(2009)\citenamefont {Megias},
  \citenamefont {Ruiz~Arriola},\ and\ \citenamefont {Salcedo}}]{Megias:2009mp}%
  \BibitemOpen
  \bibfield  {author} {\bibinfo {author} {\bibfnamefont {E.}~\bibnamefont
  {Megias}}, \bibinfo {author} {\bibfnamefont {E.}~\bibnamefont
  {Ruiz~Arriola}}, \ and\ \bibinfo {author} {\bibfnamefont {L.~L.}\
  \bibnamefont {Salcedo}},\ }\href@noop {} {\bibfield  {journal} {\bibinfo
  {journal} {Phys. Rev.},\ }\textbf {\bibinfo {volume} {D80}},\ \bibinfo
  {pages} {056005} (\bibinfo {year} {2009})}\BibitemShut {NoStop}%
\bibitem [{\citenamefont {Megias}\ \emph {et~al.}(2007)\citenamefont {Megias},
  \citenamefont {Ruiz~Arriola},\ and\ \citenamefont {Salcedo}}]{Megias:2007pq}%
  \BibitemOpen
  \bibfield  {author} {\bibinfo {author} {\bibfnamefont {E.}~\bibnamefont
  {Megias}}, \bibinfo {author} {\bibfnamefont {E.}~\bibnamefont
  {Ruiz~Arriola}}, \ and\ \bibinfo {author} {\bibfnamefont {L.}~\bibnamefont
  {Salcedo}},\ }\Doi {10.1103/PhysRevD.75.105019} {\bibfield  {journal}
  {\bibinfo  {journal} {Phys.Rev.},\ }\textbf {\bibinfo {volume} {D75}},\
  \bibinfo {pages} {105019} (\bibinfo {year} {2007})}\BibitemShut {NoStop}%
\bibitem [{\citenamefont {Riek}\ and\ \citenamefont
  {Rapp}(2010)}]{Riek:2010fk}%
  \BibitemOpen
  \bibfield  {author} {\bibinfo {author} {\bibfnamefont {F.}~\bibnamefont
  {Riek}}\ and\ \bibinfo {author} {\bibfnamefont {R.}~\bibnamefont {Rapp}},\
  }\Doi {10.1103/PhysRevC.82.035201} {\bibfield  {journal} {\bibinfo  {journal}
  {Phys. Rev.},\ }\textbf {\bibinfo {volume} {C82}},\ \bibinfo {pages} {035201}
  (\bibinfo {year} {2010})},\ \Eprint {http://arxiv.org/abs/1005.0769}
  {arXiv:1005.0769 [hep-ph]} \BibitemShut {NoStop}%
\bibitem [{\citenamefont {Mandula}\ and\ \citenamefont
  {Ogilvie}(1988)}]{Mandula:1987cp}%
  \BibitemOpen
  \bibfield  {author} {\bibinfo {author} {\bibfnamefont {J.~E.}\ \bibnamefont
  {Mandula}}\ and\ \bibinfo {author} {\bibfnamefont {M.}~\bibnamefont
  {Ogilvie}},\ }\Doi {10.1016/0370-2693(88)90091-3} {\bibfield  {journal}
  {\bibinfo  {journal} {Phys. Lett.},\ }\textbf {\bibinfo {volume} {B201}},\
  \bibinfo {pages} {117} (\bibinfo {year} {1988})}\BibitemShut {NoStop}%
\bibitem [{\citenamefont {Chernodub}\ and\ \citenamefont
  {Ilgenfritz}(2008)}]{Chernodub:2008kf}%
  \BibitemOpen
  \bibfield  {author} {\bibinfo {author} {\bibfnamefont {M.~N.}\ \bibnamefont
  {Chernodub}}\ and\ \bibinfo {author} {\bibfnamefont {E.~M.}\ \bibnamefont
  {Ilgenfritz}},\ }\href@noop {} {\bibfield  {journal} {\bibinfo  {journal}
  {Phys. Rev.},\ }\textbf {\bibinfo {volume} {D78}},\ \bibinfo {pages} {034036}
  (\bibinfo {year} {2008})}\BibitemShut {NoStop}%
\bibitem [{\citenamefont {Heinz}\ \emph {et~al.}(1987)\citenamefont {Heinz},
  \citenamefont {Kajantie},\ and\ \citenamefont {Toimela}}]{Heinz:1986kz}%
  \BibitemOpen
  \bibfield  {author} {\bibinfo {author} {\bibfnamefont {U.~W.}\ \bibnamefont
  {Heinz}}, \bibinfo {author} {\bibfnamefont {K.}~\bibnamefont {Kajantie}}, \
  and\ \bibinfo {author} {\bibfnamefont {T.}~\bibnamefont {Toimela}},\ }\Doi
  {10.1016/0003-4916(87)90002-9} {\bibfield  {journal} {\bibinfo  {journal}
  {Ann. Phys.},\ }\textbf {\bibinfo {volume} {176}},\ \bibinfo {pages} {218}
  (\bibinfo {year} {1987})}\BibitemShut {NoStop}%
\bibitem [{\citenamefont {Weldon}(1982)}]{Weldon:1982aq}%
  \BibitemOpen
  \bibfield  {author} {\bibinfo {author} {\bibfnamefont {H.}~\bibnamefont
  {Weldon}},\ }\Doi {10.1103/PhysRevD.26.1394} {\bibfield  {journal} {\bibinfo
  {journal} {Phys.Rev.},\ }\textbf {\bibinfo {volume} {D26}},\ \bibinfo {pages}
  {1394} (\bibinfo {year} {1982})}\BibitemShut {NoStop}%
\bibitem [{\citenamefont {Reinders}\ \emph {et~al.}(1985)\citenamefont
  {Reinders}, \citenamefont {Rubinstein},\ and\ \citenamefont
  {Yazaki}}]{Reinders:1984sr}%
  \BibitemOpen
  \bibfield  {author} {\bibinfo {author} {\bibfnamefont {L.}~\bibnamefont
  {Reinders}}, \bibinfo {author} {\bibfnamefont {H.}~\bibnamefont
  {Rubinstein}}, \ and\ \bibinfo {author} {\bibfnamefont {S.}~\bibnamefont
  {Yazaki}},\ }\Doi {10.1016/0370-1573(85)90065-1} {\bibfield  {journal}
  {\bibinfo  {journal} {Phys. Rept.},\ }\textbf {\bibinfo {volume} {127}},\
  \bibinfo {pages} {1} (\bibinfo {year} {1985})}\BibitemShut {NoStop}%
\bibitem [{\citenamefont {Chakraborty}\ \emph {et~al.}()\citenamefont
  {Chakraborty}, \citenamefont {Mustafa},\ and\ \citenamefont
  {Thoma}}]{cmt:2011b}%
  \BibitemOpen
  \bibfield  {author} {\bibinfo {author} {\bibfnamefont {P.}~\bibnamefont
  {Chakraborty}}, \bibinfo {author} {\bibfnamefont {M.~G.}\ \bibnamefont
  {Mustafa}}, \ and\ \bibinfo {author} {\bibfnamefont {M.~H.}\ \bibnamefont
  {Thoma}},\ }\href@noop {} {}\bibinfo {note} {In preparation}\BibitemShut
  {NoStop}%
\end{thebibliography}%

\end{document}